\documentclass[sigconf]{acmart}

\pdfoutput=1

\renewcommand\footnotetextcopyrightpermission[1]{} 
\makeatletter
\renewcommand\@formatdoi[1]{\ignorespaces}
\makeatother

\usepackage{silence}
\usepackage{amssymb,amsmath,amsthm}
\usepackage{graphicx}
\usepackage{cleveref}[2012/02/15]
\usepackage{mathtools}																
\usepackage{soul}																			
\usepackage{url}
\usepackage{color, colortbl}
\usepackage{xifthen}
\usepackage{xparse}
\usepackage{xstring}                                  
\usepackage{paralist}                                 
\usepackage{stmaryrd}																	
\usepackage{floatflt}                                 
\usepackage{float}                                    
\usepackage[caption=false]{subfig}
\usepackage{mfirstuc}                                 
\usepackage{array}
\usepackage{balance}
\usepackage{textcomp}                                 
\usepackage{comment}
\usepackage{multirow}
\usepackage{pifont}
\usepackage{enumitem}
\usepackage{flushend}                                 
 
\setlist[enumerate]{leftmargin=*,nosep}
\setlist[itemize]{leftmargin=*,nosep}

\definecolor{CommentColor}{RGB}{128,128,128}
\usepackage[ruled,vlined,linesnumbered]{algorithm2e}
\SetKwInput{KwData}{Input}
\SetKwInput{KwResult}{Output}

\SetCommentSty{MyCommFont}
\SetKwProg{Function}{Function}{}{}
\DontPrintSemicolon

\crefformat{footnote}{#2\footnotemark[#1]#3}

\hypersetup{
  pdftitle={Revisiting MITL: Impossible Semantics, Incorrect Decision Procedures},
  pdfauthor={Nima Roohi, Mahesh Viswanathan},
  colorlinks=true,
  citecolor={RoyalPurple},
  linkcolor = {violet},
  bookmarksopen=false,
  bookmarksnumbered=true
}

\definecolor{redish}      {rgb}{0.8, 0.1, 0.1}
\definecolor{blueish}     {rgb}{0  , 0  , 0.4}
\definecolor{greenish}    {rgb}{0  , 0.6, 0  }
\definecolor{yellowish}   {rgb}{0.8, 0.5, 0  }
\definecolor{redishBG}    {rgb}{1  , 0.3, 0.3}
\definecolor{blueishBG}   {rgb}{0.4, 0.4, 1  }
\definecolor{greenishBG}  {rgb}{0.2, 1  , 0.2}
\definecolor{yellowishBG} {rgb}{1  , 0.7, 0  }

\newtheoremstyle{red-th}{}{}{}{}{\color{redish}\bfseries}{.}{ }{} 
\newtheoremstyle{blue-th}{}{}{}{}{\color{blueish}\bfseries}{.}{ }{} 
\newtheoremstyle{green-th}{}{0em}{}{}{\color{greenish}\bfseries}{.}{ }{} 

\newcommand{\Asm}   {Assumption}
\newcommand{\Alg}   {Algorithm}
\newcommand{\Apx}   {Appendix}
\newcommand{\Cor}   {Corollary}
\newcommand{\Def}   {Definition}
\newcommand{\Eq}    {Equation}
\newcommand{\Fig}   {Figure}
\newcommand{\Lem}   {Lemma}
\newcommand{\Prop}  {Proposition}
\newcommand{\Rmrk}  {Remark}
\newcommand{\Sec}   {Section}
\newcommand{\Tbl}   {Table}
\newcommand{\Thm}   {Theorem}
\newcommand{\Cnd}   {Condition}
\newcommand{\Prb}   {Problem}
\newcommand{\Ex}    {Example}
\newcommand{\Itm}   {Item}
\newcommand{\Line}  {Line}
\newcommand{\Fml}   {Formula}

\NewDocumentCommand{\aref}{m}{%
  \IfBeginWith{#1}{lin:} {\Line~\ref{#1}}{%
  \IfBeginWith{#1}{asm:} {\Asm~\ref{#1}}{%
  \IfBeginWith{#1}{sec:} {\Sec~\ref{#1}}{%
  \IfBeginWith{#1}{thm:} {\Thm~\ref{#1}}{%
  \IfBeginWith{#1}{lem:} {\Lem~\ref{#1}}{%
  \IfBeginWith{#1}{cor:} {\Cor~\ref{#1}}{%
  \IfBeginWith{#1}{def:} {\Def~\ref{#1}}{%
  \IfBeginWith{#1}{apx:} {\Apx~\ref{#1}}{%
  \IfBeginWith{#1}{alg:} {\Alg~\ref{#1}}{%
  \IfBeginWith{#1}{fig:} {\Fig~\ref{#1}}{%
  \IfBeginWith{#1}{tbl:} {\Tbl~\ref{#1}}{%
  \IfBeginWith{#1}{rem:} {\Rem~\ref{#1}}{%
  \IfBeginWith{#1}{cnd:} {\Cnd~\ref{#1}}{%
  \IfBeginWith{#1}{prb:} {\Prb~\ref{#1}}{%
  \IfBeginWith{#1}{itm:} {\Itm~\ref{#1}}{%
  \IfBeginWith{#1}{eq:}  {\Eq~\ref{#1}}{%
  \IfBeginWith{#1}{ex:}  {\Ex~\ref{#1}}{%
  \IfBeginWith{#1}{fml:} {\Fml~\ref{#1}}{%
  \IfBeginWith{#1}{prop:}{\Prop~\ref{#1}}{%
  \errmessage{class of label '#1' is not defined.}%
  }}}}}}}}}}}}}}}}}}}%
}

\newtheorem{theorem}             {\Thm}
\newtheorem{definition}  [theorem] {\Def}                

\newtheorem{problem}   [theorem] {\Prb}

\makeatletter
\newtheorem*{rep@theorem}{\rep@title}
\newcommand{\newreptheorem}[2]{%
\newenvironment{rep#1}[1]{%
 \def\rep@title{#2 \ref{##1}}%
 \begin{rep@theorem}}%
 {\end{rep@theorem}}}
\makeatother

\newreptheorem{corollary}   {\Corl}                         
\newreptheorem{lemma}       {\Lem}
\newreptheorem{theorem}     {\Thm}
\newreptheorem{proposition} {\Prop}

\DeclarePairedDelimiter{\Size}    {\vert}{\vert}
\DeclarePairedDelimiter{\Paren}   {\lparen}{\rparen}
\DeclarePairedDelimiter{\Brace}   {\lbrace}{\rbrace}

\DeclarePairedDelimiter{\LclRop}  {\lbrack}{\rparen}
\DeclarePairedDelimiter{\LopRcl}  {\lparen}{\rbrack}
\DeclarePairedDelimiter{\Closed}  {\lbrack}{\rbrack}
\DeclarePairedDelimiter{\Opened}  {\lparen}{\rparen}

\DeclareExpandableDocumentCommand{\LeftQ} {}{\scalebox{-0.7}[0.7]{\raisebox{-0.5mm}{?}}\!}
\DeclareExpandableDocumentCommand{\RightQ}{}{\!\scalebox{0.7}[0.7]{\raisebox{-0.5mm}{?}}}

\NewDocumentCommand{\GEz}     {}{{\scalebox{0.5}{${\geq}0$}}}
\NewDocumentCommand{\GTz}     {}{{\scalebox{0.6}{$+$}}}

\NewDocumentCommand{\Nat}     {}{\ensuremath{\mathbb{N}}}

\NewDocumentCommand{\Real}    {}{\ensuremath{\mathbb{R}}}
\NewDocumentCommand{\pNat}    {}{\ensuremath{\Nat_\GTz}}

\NewDocumentCommand{\pReal}   {}{\ensuremath{\Real_\GTz}}

\NewDocumentCommand{\nnReal}  {}{\ensuremath{\Real_\GEz}}

\NewDocumentCommand{\iFF} {}{\mbox{iff}}    

\NewDocumentCommand{\ie}  {}{{\em i.e.}}

\NewDocumentCommand{\aka} {}{{\em a.k.a.}}

\DeclareExpandableDocumentCommand{\BNF}   {}{\texttt{BNF}}
\DeclareExpandableDocumentCommand{\NNF}   {}{\texttt{NNF}}
\DeclareExpandableDocumentCommand{\CNF}   {}{\texttt{CNF}}
\DeclareExpandableDocumentCommand{\DNF}   {}{\texttt{DNF}}
\DeclareExpandableDocumentCommand{\MTL}   {}{\texttt{MTL}}
\DeclareExpandableDocumentCommand{\LTL}   {}{\texttt{LTL}}
\DeclareExpandableDocumentCommand{\MITL}  {}{\texttt{MITL}}
\DeclareExpandableDocumentCommand{\MITLzi}{}{\MITL\textsubscript{$0,\!\infty$}}
\DeclareExpandableDocumentCommand{\MITLwi}{}{\MITL\textsubscript{WI}}
\DeclareExpandableDocumentCommand{\STL}   {}{\texttt{STL}}
\DeclareExpandableDocumentCommand{\prSTL} {}{\texttt{prSTL}}
\DeclareExpandableDocumentCommand{\PAIN}  {}{\texttt{PAIN}}
 
\NewDocumentCommand{\PVS} {}{\ensuremath{\mathtt{PVS}}}
\NewDocumentCommand{\Lean}{}{\ensuremath{\mathtt{Lean}}}

\NewDocumentCommand{\PVSLabel}{smm}{%
\scriptsize%
\texttt{#3}@\texttt{#2}\IfBooleanF{#1}{.}%
}

\NewDocumentCommand{\WeHave}  {}{\scalebox{0.3}{\ }\raisebox{-1ex}{\scalebox{2.3}{$\mathrel{\cdot}$}}}
\NewDocumentCommand{\SuchThat}{}{\WeHave}

\NewDocumentCommand{\oftype}  {}{\ensuremath{\mathrel{:}}}

\NewDocumentCommand{\goesto}  {O{}} {\ensuremath{\xrightarrow{#1}}}
\NewDocumentCommand{\Implies} {O{}} {\ensuremath{\xrightarrow{#1}}}
 
\NewDocumentCommand{\AAA}{}{\ensuremath{\mathcal{A}}}

\NewDocumentCommand{\CCC}{}{\ensuremath{\mathcal{C}}}

\NewDocumentCommand{\OOO}{}{\ensuremath{\mathcal{O}}}

\NewDocumentCommand{\Val}   {}{\ensuremath{\nu}}

\NewDocumentCommand{\defEQ}     {}   {\ensuremath{\coloneqq}}
\NewDocumentCommand{\Sat}       {}   {\ensuremath{\models}}

\NewDocumentCommand{\cart}			{}   {\ensuremath{\times}}

\definecolor{DarkGray}{RGB}{70,70,70}
\DeclareExpandableDocumentCommand{\GrayColor}{}{DarkGray}

\DeclareExpandableDocumentCommand{\CONrawstr}{}{\mathtt{CNT}}
\DeclareExpandableDocumentCommand{\DISrawstr}{}{\mathtt{DSC}}

\DeclareExpandableDocumentCommand{\CONstr}{}{\scalebox{0.4}{\ensuremath{\color{\GrayColor}\CONrawstr}}}
\DeclareExpandableDocumentCommand{\DISstr}{}{\scalebox{0.4}{\ensuremath{\color{\GrayColor}\DISrawstr}}}
\newlength{\SatLength}
\newlength{\SatHeight}
\NewDocumentCommand{\Dom}    {o} {\ensuremath{\mathtt{dom}\IfValueT{#1}{\Paren{#1}}}}

\makeatletter
\newcommand{\Minos}{\mathbin{\text{\@Minos}}}
\newcommand{\@Minos}{%
  \ooalign{\hidewidth\raise1ex\hbox{.}\hidewidth\cr$\m@th-$\cr}%
}
\makeatother

\DeclareExpandableDocumentCommand{\NextOp}      {}{\ensuremath{\scalebox{0.8}{$\bigcirc$}}}
\DeclareExpandableDocumentCommand{\EventuallyOp}{}{\ensuremath{\lozenge}}
\DeclareExpandableDocumentCommand{\AlwaysOp}    {}{\ensuremath{\square}}
\DeclareExpandableDocumentCommand{\UntilOp}     {}{\ensuremath{\mathcal{U}}}
\DeclareExpandableDocumentCommand{\ReleaseOp}   {}{\ensuremath{\mathcal{R}}}
\NewDocumentCommand{\Until}     {mmo}{\ensuremath{#1\UntilOp\IfValueT{#3}{_{#3}}#2}}
\NewDocumentCommand{\Release}   {mmo}{\ensuremath{#1\ReleaseOp\IfValueT{#3}{_{#3}}#2}}
\NewDocumentCommand{\Next}      {mo} {\ensuremath{\NextOp\IfValueT{#2}{_{#2}}#1}}
\NewDocumentCommand{\Always}    {mo} {\ensuremath{\AlwaysOp\IfValueT{#2}{_{#2}}#1}}
\NewDocumentCommand{\Eventually}{mo} {\ensuremath{\EventuallyOp\IfValueT{#2}{_{#2}}#1}}

\NewDocumentCommand{\Obs}{o}{\ensuremath{\OOO\IfValueT{#1}{\Paren{#1}}}}
\NewDocumentCommand{\Obss}{mo}{\ensuremath{\OOO_#1\IfValueT{#2}{\Paren{#2}}}}

\NewDocumentCommand{\UnaryFunc}{mmo}{\ensuremath{%
  \IfValueTF{#3}%
  {\IfBooleanTF{#1}%
    {#2\Paren*{#3}}%
    {#2\Paren{#3}}%
  }%
  {#2}%
}}

\NewDocumentCommand{\FUNCTION}{mm}{%
  \expandafter\NewDocumentCommand\expandafter{\csname #1\endcsname}{so}{%
    #2%
    \IfNoValueF{##2}{\IfBooleanTF{##1}{\Paren*{##2}}{\Paren{##2}}}
  }%
}

\NewDocumentCommand{\FunctionWithSub}{mm}{%
  \expandafter\NewDocumentCommand\expandafter{\csname #1\endcsname}{smo}{%
		#2_{##2}%
    \IfNoValueF{##3}{\IfBooleanTF{##1}{\Paren*{##3}}{\Paren{##3}}}
  }%
}

\NewDocumentCommand{\FunctionWithSup}{mm}{%
  \expandafter\NewDocumentCommand\expandafter{\csname #1\endcsname}{smo}{%
		#2^{##2}%
    \IfNoValueF{##3}{\IfBooleanTF{##1}{\Paren*{##3}}{\Paren{##3}}}
  }%
}

\NewDocumentCommand{\FunctionWithBoth}{mm}{%
  \expandafter\NewDocumentCommand\expandafter{\csname #1\endcsname}{smmo}{%
		#2^{##2}_{##3}%
    \IfValueT{##4}{\IfBooleanTF{##1}{\Paren*{##4}}{\Paren{##4}}}
  }%
}

\NewDocumentCommand{\fvarR}{so}{\UnaryFunc{#1}{\mathtt{fvar_R}}[#2]}
\NewDocumentCommand{\fvarL}{so}{\UnaryFunc{#1}{\mathtt{fvar_L}}[#2]}
\NewDocumentCommand{\fvar}{so}{\UnaryFunc{#1}{\mathtt{fvar}}[#2]}

\NewDocumentCommand{\nnf}  {so}{\UnaryFunc{#1}{\mathtt{nnf}}[#2]}
\NewDocumentCommand{\toOld}{so}{\UnaryFunc{#1}{\mathtt{old}}[#2]}
\NewDocumentCommand{\Depth}{so}{\UnaryFunc{#1}{\mathtt{depth}}[#2]}
\NewDocumentCommand{\Step} {so}{\UnaryFunc{#1}{\mathtt{sz}}[#2]}
\NewDocumentCommand{\DS}   {so}{\UnaryFunc{#1}{\mathtt{DS}}[#2]}

\NewDocumentCommand{\FTypeP}{mm}{\ensuremath{{#2}^{#1}}}
\NewDocumentCommand{\FTypeA}{mm}{\ensuremath{{#1}\goesto{#2}}}

\FUNCTION{HDist}  {\mathtt{d}_H}
\FUNCTION{infDist}{\mathtt{d}_\infty}
\FUNCTION{infBall}{\mathtt{B}_\infty}
\FUNCTION{Traj}   {\mathtt{Traj}}
\FUNCTION{Exec}		{\mathtt{Exec}}
\FUNCTION{Closure}{\mathtt{cl}}
\FUNCTION{Adist}  {\infDist^\rightarrow}
\FUNCTION{Var}    {\mathtt{fvar}}

\NewDocumentCommand{\Term}{mmm}{%
  \expandafter\DeclareExpandableDocumentCommand\expandafter{\csname #1ls\endcsname}{}{\MakeLowercase{#2}}%
  \expandafter\DeclareExpandableDocumentCommand\expandafter{\csname #1lp\endcsname}{}{\MakeLowercase{#3}}%
  \expandafter\DeclareExpandableDocumentCommand\expandafter{\csname #1cs\endcsname}{}{#2}%
  \expandafter\DeclareExpandableDocumentCommand\expandafter{\csname #1cp\endcsname}{}{#3}%
  \expandafter\DeclareExpandableDocumentCommand\expandafter{\csname #1us\endcsname}{}{\makefirstuc{\MakeLowercase{#2}}}%
  \expandafter\DeclareExpandableDocumentCommand\expandafter{\csname #1up\endcsname}{}{\makefirstuc{\MakeLowercase{#3}}}%
} 

\NewDocumentCommand{\STORMED}{}{\mbox{\scalebox{0.9}{STORMED}}}
\NewDocumentCommand{\Buchi}{}{B\"uchi}

\Term{BA}       {\Buchi\ Automaton}                      {\Buchi\ Automata}
\Term{TS}       {Transition System}                      {Transition Systems}
\Term{HS}       {Hybrid System}													 {Hybrid Systems}
\Term{HA}       {Hybrid Automaton}                       {Hybrid Automata}
\Term{TA}       {Timed Automaton}                        {Timed Automata}
\Term{rTA}      {Rational Timed Automaton}               {Rational Timed Automata}
\Term{irTA}     {Irrational Timed Automaton}             {Irrational Timed Automata}
\Term{PolyHA}   {Polyhedral Automaton}                   {Polyhedral Automata}
\Term{nlfPolyHA}{Non-linear Polyhedral Automaton}        {Non-linear Polyhedral Automata}
\Term{RHA}      {Rectangular Automaton}                  {Rectangular Automata}
\Term{MRHA}     {Monotonic Rectangular Automaton}        {Monotonic Rectangular Automata}
\Term{iMRHA}    {Initialzied Monotonic Rectangular Automaton}        {Initialized Monotonic Rectangular Automata}
\Term{LIHA}     {Linear Inclusion  Automaton}            {Linear Inclusion Automata}
\Term{iRHA}     {Initialized Rectangular Automaton}      {Initialized Rectangular Automata}
\Term{iLIHA}    {Initialized Linear Inclusion Automaton} {Initialized Linear Inclusion Automata}
\Term{AHA}      {Affine Hybrid Automaton}                {Affine Hybrid Automata}
\Term{LEHA}     {Linear Equation Automaton}              {Linear Equation Automata}
\Term{iLEHA}    {Initialized Linear Equation Automaton}  {Initialized Linear Equation Automata}
\Term{SWHA}     {Stopwatch Automaton}                    {Stopwatch Automata}
\Term{iSWHA}    {Initialized Stopwatch Automaton}        {Initialized Stopwatch Automata}
\Term{SHA}      {Solvable Hybrid Automaton}              {Solvable Hybrid Automata}
\Term{SAOS}     {Semi-Algebraic O-Minimal System}        {Semi-Algebraic O-Minimal Systems}
\Term{SASS}			{Semi-Algebraic \STORMED\ System}			   {Semi-Algebraic \STORMED\ Systems}
\Term{tCM}			{$2$-Counter Machine}									   {$2$-Counter Machines}

\NewDocumentCommand{\aMaxAccl} {}{\ensuremath{a_{\texttt{max\_accel}}}}
\NewDocumentCommand{\aMinBrake}{}{\ensuremath{a_{\texttt{min\_brake}}}}
\NewDocumentCommand{\aMaxBrake}{}{\ensuremath{a_{\texttt{max\_brake}}}}

\NewDocumentCommand{\dMinLon}     {}{\ensuremath{d^{\texttt{lon}}_{\texttt{min}}}}        
\NewDocumentCommand{\DangLon}     {}{\ensuremath{\textsf{dang\textsuperscript{\texttt{lon}}}}}
\NewDocumentCommand{\BlameLon}    {}{\ensuremath{\textsf{blame\textsuperscript{\texttt{lon}}}}}

\NewDocumentCommand{\aMaxAcclLat} {}{\ensuremath{a^{\texttt{lat}}_{\texttt{max\_accel}}}}
\NewDocumentCommand{\aMaxBrakeLat}{}{\ensuremath{a^{\texttt{lat}}_{\texttt{max\_brake}}}}
\NewDocumentCommand{\aMinBrakeLat}{}{\ensuremath{a^{\texttt{lat}}_{\texttt{min\_brake}}}}
\NewDocumentCommand{\dMinLat}     {}{\ensuremath{d^{\texttt{lat}}_{\texttt{min}}}}        
\NewDocumentCommand{\DangLat}     {}{\ensuremath{\textsf{dang\textsuperscript{\texttt{lat}}}}}
\NewDocumentCommand{\BlameLat}    {}{\ensuremath{\textsf{blame\textsuperscript{\texttt{lat}}}}}

\NewDocumentCommand{\Dang}     {}{\ensuremath{\textsf{dang}}}
\NewDocumentCommand{\Blame}    {}{\ensuremath{\textsf{blame}}}

\NewDocumentCommand{\Policy}   {}{\ensuremath{\mathbb{P}}}

\NewDocumentCommand{\POS}      {}{\ensuremath{\texttt{Pos}}}
\NewDocumentCommand{\LOC}      {}{\ensuremath{\texttt{Loc}}}
\NewDocumentCommand{\DELAY}    {}{\ensuremath{\texttt{Delay}}}

\NewDocumentCommand{\dReach}     {}{\texttt{dReach}}
\NewDocumentCommand{\SpaceEx}    {}{\texttt{SpaceEx}}
\NewDocumentCommand{\PHAVer}     {}{\texttt{PHAVer}}
\NewDocumentCommand{\HyTech}     {}{\texttt{HyTech}}
\NewDocumentCommand{\CtEt}       {}{\texttt{C2E2}}
\NewDocumentCommand{\FlowS}      {}{\texttt{Flow\textsuperscript{*}}}
\NewDocumentCommand{\HARE}       {}{\texttt{HARE}}
\NewDocumentCommand{\HSolver}    {}{\texttt{HSolver}}
\NewDocumentCommand{\UPPALL}     {}{\texttt{UPPAAL}}

\title{Self-Driving Vehicle Verification Towards a Benchmark}
    
\author{Nima Roohi}
\affiliation{%
  \institution{University of Pennsylvania}
  \streetaddress{3330 Walnut St.}
  \city{Philadelphia} 
  \state{Pennsylvania} 
  \postcode{19104}
}
\email{roohi2@cis.upenn.edu}

\author{Ramneet Kaur}
\affiliation{%
  \institution{University of Pennsylvania}
  \streetaddress{3330 Walnut St.}
  \city{Philadelphia} 
  \state{Pennsylvania} 
  \postcode{19104}
}
\email{ramneetk@seas.upenn.edu}

\author{James Weimer}
\affiliation{%
  \institution{University of Pennsylvania}
  \streetaddress{3330 Walnut St.}
  \city{Philadelphia} 
  \state{Pennsylvania} 
  \postcode{19104}
}
\email{weimerj@seas.upenn.edu}
 
\author{Oleg Sokolsky}
\affiliation{%
  \institution{University of Pennsylvania}
  \streetaddress{3330 Walnut St.}
  \city{Philadelphia} 
  \state{Pennsylvania} 
  \postcode{19104}
}
\email{sokolsky@cis.upenn.edu}

\author{Insup Lee}
\affiliation{%
  \institution{University of Pennsylvania}
  \streetaddress{3330 Walnut St.}
  \city{Philadelphia} 
  \state{Pennsylvania} 
  \postcode{19104}
}
\email{lee@cis.upenn.edu}
   
\begin{document}
\begin{abstract}
Industrial cyber-physical systems are hybrid systems with strict safety requirements.
Despite not having a formal semantics, most of these systems are modeled using Stateflow/Simulink\textsuperscript\textregistered\ for mainly two reasons: 
\begin{inparaenum}
\item it is easier to model, test, and simulate using these tools, and
\item dynamics of these systems are not supported by most other tools.
\end{inparaenum}
Furthermore, 
with the ever growing complexity of cyber-physical systems, grows the gap between what can be modeled using an automatic {\em formal} verification tool and 
models of industrial cyber-physical systems.
In this paper, we present a simple formal model for self-deriving cars.
While after some simplification, safety of this system has already been proven manually, to the best of our knowledge, no automatic formal verification tool 
supports its dynamics.
We hope this serves as a challenge problem for formal verification tools targeting industrial applications. 
\end{abstract}



\maketitle  

\section{Introduction}\label{sec:intro}
The following paragraph is taken directly from~\cite{17-mob}.
According to the authors, two ingredients are missing from the race for who will have the first self-driving car on the road:
\begin{inparaenum}
\item standardization of safety assurance, and
\item scalability.
\end{inparaenum}
\begin{quote}
The ``Winter of AI'' is commonly known as the decades long period of inactivity following the collapse of Artificial Intelligence research that over-reached 
its goals and hyped its promise until the inevitable fall during the early 80s. We believe that the development of Autonomous Vehicles (AV) is dangerously 
moving along a similar path that might end in great disappointment after which further progress will come to a halt for many years to come.
\end{quote}

A typical approach to estimate the amount of safety assurance while preserving scalability, is to use statistical techniques, in which one simulates the system 
or collects actual/random data.
To appreciate the problematic nature of a data-driven approach, authors in~\cite{17-mob} prove, in order to have $10^{-9}$ fatality per hour in an autonomous 
vehicle, one require $10^9$ hours of data, which for example means, 1000 employees must drive 24 hours a day, 7 days a week, for 114 years!
Even worse, every time part of a system gets updated, no matter how small, preserving the guarantee requires repeating the whole data collection.

To solve the safety standardization problem, \cite{17-mob} suggests the notion of ``who is responsible'' for an accident in a non-deterministic setting.
Intuitively, instead of trying to build a system in which no accident occurs, ``Responsibility-Sensitive Safety'', tries to prevent car $c$ from only those
accidents in which $c$ is going to be blamed. In other words, if a car drives responsibly, it might still be involved in an accident, but it will never be
blamed for one.
To achieve this goal, \cite{17-mob} defines two major components:
\begin{inparaenum}
\item a policy that cars should follow, and
\item a mechanism to identify responsible parity (or parties) in case of an accident (they are exactly those who will be blamed for the accident). 
\end{inparaenum}
The majority of the paper is devoted to different policies in different conditions, like
  moving in the same direction or in the opposite directions,
  moving laterally or longitudinally,
  moving on a straight road or road with other geometries,
and who should be blamed in case of accident in each of those conditions.

After a policy is defined, one has to show it does not blame those who follow it and at least one party in each accident will be blamed (otherwise a policy 
that prevents nothing won't blame anyone for an accident, but also does not prevent any accident from happening).
Unfortunately, all these proofs or in some cases only sketches of proofs are done manually in~\cite{17-mob}.
However, to the best of our knowledge, there is no automatic formal verification tool that can be used to prove these properties.
Even worse, we are not aware of any automatic formal verification tool that can be used to specify these properties. 
This was our motivation to write this paper,
in which we specify the most basic and fundamental policies defined in~\cite{17-mob} and challenge current and any future automatic formal verification tool for
cyber-physical system to solve any of the five challenge problems in this paper.

In \aref{sec:prelim}, we review preliminary definitions we need in this paper.
In \aref{sec:spec}, we formally specify system and policy defined in~\cite{17-mob}, for the case when finite number of cars are driving on a straight road. 
The rigorous level of the specifications in this paper is high enough to seamlessly write them all in a theorem prover like 
\PVS~\cite{92-PVS}\ or \Lean~\cite{15-Lean}.
This removes any ambiguities from policy and system dynamics~\footnote{Through this process we observed a couple of problems/inconsistencies with the 
specifications in~\cite{17-mob} (they are mentioned at different places in this paper).}.
Having policy and system dynamics clearly defined, next we specify five different fundamental problems about these specification in \aref{sec:probs}.
The first four are about (robust) safety and (robust) liveness, and the last one, is about the policy when not every car follows it.
We use signals (a function from a non-negative real value as time to a point in a metric space) to specify all of our system dynamics, policy, and problems.
This makes our specifications uniform but not constructive, \ie\ it does not specify how to build a system that follows those specifications.
In \aref{sec:tools}, we list nine different tools and six different reasons that prevent us from even specifying our problems in these tools.
This is after ignoring all the difficulties that may arise when one wants to encode everything in the language of one of these tools.
All these tools are written solely for the purpose of formal model checking cyber-physical systems.
Finally, we conclude the paper in \aref{sec:conc}.

\section{Preliminaries}\label{sec:prelim}
We denote the set of {\em natural}, {\em positive natural}, {\em real}, {\em positive real}, and {\em non-negative real} numbers by
                     $\Nat$,        $\pNat$,                $\Real$,    $\pReal$,            and $\nnReal$, respectively.
For any two sets $A$ and $B$,
  {\em size} of $A$ is denoted by $\Size{A}$,
and 
  {\em the set of functions from $A$ to $B$} is denoted by $\FTypeA{A}{B}$ or $\FTypeP{A}{B}$.
Operator $\goesto$ is considered to be right-associative, meaning if $C$ is a set then function $f$ of type $A\goesto B\goesto C$
is a function that maps every element of type $A$ to an element of type $B \goesto C$.

\subsection{Extended Metric Space and Distance Functions}
Let $M$ be an arbitrary set and $d\oftype\FTypeA{M\cart M}{\Real\cup\Brace{\infty}}$ be an arbitrary function.
Ordered pair $(M,d)$ is called an {\em extended metric space} and $d$ is called a {\em distance function} \iFF\ 
for any $x,y,z\oftype M$ the following conditions hold:
\begin{inparaenum}
\item $d(x,y)\geq0$, 
\item $d(x,y)=0\Leftrightarrow x=y$,
\item $d(x,y)=d(y,x)$, and
\item $d(x,z)\leq d(x,y)+d(y,z)$.
\end{inparaenum}
If distance function $d$ is known from the context, we use $M$ to also refer to the metric space.
Let $X$ be a finite set of variables, and $M\subseteq\FTypeP{X}{\Real}$ be an arbitrary set.
A well-known distance function on $M$, denoted by $\infDist[\Val_1,\Val_2]$, maps any two points $\Val_1,\Val_2\oftype M$ to 
$\max\limits_{x\oftype X}\Size{\Val_1(x)-\Val_2(x)}$.
Let $C$ be a finite set and $(M,d)$ be a metric space.
We extend $d$ to map any two points $\Val_1,\Val_2\oftype\FTypeA{C}{M}$ to 
$\max\limits_{c\oftype C}d\Paren*{\Val_1(c),\Val_2(c)}$.
\subsection{Signal}
In this paper we present dynamics of a system and policies both using signals.
This is for two reasons:
First, using one formalism to specify both behavior and policy makes our presentation more uniform.
Second, we intentionally stay away from any class of hybrid automata or temporal logic, and leave it to the reader to choose or develop an appropriate formalism
for this benchmark.

\begin{definition}[Signal]\label{def:signal}
Let $(M,d)$ be an extended metric space.
Signal is any function of type $\FTypeA{\nnReal}{M}$.
\end{definition}

Signal $f$ is {\em continuous at time $t\oftype\nnReal$} \iFF\ 
$\lim_{t'\goesto t}f(t')$ is defined and equal to $f(t)$~\footnote{When $t=0$, we only consider continuity from right.}.
Signal $f$ is {\em continuous} \iFF\ it is continuous at all times.
Signal $f$ is {\em piecewise continuous} \iFF\ number of discontinuities within any finite amount of time is finite.

\section{Dynamics and Policy}\label{sec:spec}
 
In this paper we only consider the case in which cars are on a straight road and no car drives in reverse gear.
While making the presentation simpler, this is quite enough to specify our automatic formal verification challenge problems.
\aref{tbl:params} lists every parameter that we use in this paper.

\begin{table}[H]
\renewcommand{\arraystretch}{1.2}
\NewDocumentCommand{\MAX}{}{Maximum}
\NewDocumentCommand{\MIN}{}{\makebox[0pt][l]{Minimum}\phantom{\MAX}}
\centering
\begin{tabular}{|l|l|}
\rowcolor{black}
\color{white}Parameter    & \color{white}Description \\
\CCC                      & Finite Set of Cars        \\\hline
$\rho\oftype\nnReal$      & Response Time             \\\hline
$\mu \oftype\pReal$       & Minimum Distance Parameter\\\hline
$\aMaxAccl      \oftype\pReal$ & \MAX\ Longitudinal Acceleration\\\hline
$\aMinBrake     \oftype\pReal$ & \MIN\ Longitudinal Deceleration\\\hline
$\aMaxBrake     \oftype\pReal$ & \MAX\ Longitudinal Deceleration\\\hline
$\aMaxAcclLat   \oftype\pReal$ & \MAX\ Lateral      Acceleration\\\hline
$\aMinBrakeLat  \oftype\pReal$ & \MIN\ Lateral      Deceleration\\\hline
$\aMaxBrakeLat  \oftype\pReal$ & \MAX\ Lateral      Deceleration\\\hline
\end{tabular}
\caption{Parameters}\label{tbl:params}
\end{table}

The most basic signal in this paper is the position signal that specifies position of every car on the road throughout the entire time.
The next most fundamental signal in this paper is the delay signal that models delays in the controller parts of a cyber-physical system.
We first define position and delay signals.
Next, we define minimum longitudinal and lateral distance signals as a function of position and delay signals.
Later, we use these four signals to uniquely define dangerous situation and blame time signals.
These signals together are almost everything we need to define policy and verification problems about that policy.

\begin{definition}[Position Signal]\label{def:pos-sig}
Let $\CCC$ be an arbitrary finite set of cars.
{\em Position signal} is any function of the type
$f\oftype\FTypeA{\nnReal}{\FTypeA{\CCC}{\FTypeP{\Brace{x,y}}{\Real}}}$.
\end{definition}

Note that in \aref{def:pos-sig}, $\FTypeA{\CCC}{\FTypeP{\Brace{x,y}}{\Real}}$ is the metric space.
Also, in \aref{def:pos-sig} and every other signal that is defined later in this paper, we use $\infDist$ as the distance function.
Let $f$ be a position signal.
We say $f$ is {\em differentiable at time $t\oftype\nnReal$} \iFF\
for every car $c\oftype\CCC$ and axis $u\oftype\Brace{x,y}$,
$\lim_{h\goesto0}\frac{(f\ (t+h)\ c\ u)-(f\ t\ c\ u)}{h}$ is defined~\footnote{%
  \label{fnt:at0}Similar to the continuity definition, if $t=0$, we only consider differentiability from right.}.
We say $f$ is {\em differentiable} \iFF\ it is differentiable at all times in $\nnReal$.
Furthermore, we use $f_v$ to denote derivative of $f$ and call it {\em velocity signal}
(note that $f_v$ is also a position signal).
Furthermore,
if $f_v$ is differentiable, we denote the induced derivative signal by $f_a$ and call it {\em acceleration signal}.
Finally, we use $\POS_\CCC$ to denote the set of position signals $f$ with two conditions: 
\begin{inparaenum}
\item both $f_v$ and $f_a$ are defined throughout the entire time domain, and 
\item no car has a negative longitudinal velocity (\ie\ $\forall t\oftype\nnReal,c\oftype\CCC\WeHave f_v\ t\ c\ y \geq0$).
\end{inparaenum}


Position of different cars is a physical property of our cyber-physical system and for every signal, each car has a unique position at every single point in 
time. However, when a car uses its sensors to observe positions of different cars including itself, there are at least two sources of errors:
\begin{inparaenum}
\item {\em measurement errors} caused by inaccuracy of sensors, and 
\item slight {\em delay} in sensors and controllers (parameter $\rho$ in \aref{tbl:params}).
\end{inparaenum}
To handle measurement errors, one has to consider not only a position signal, but {\em at least} all the position signals that are pointwise close to it.
This usually happens in the context of {\em robust verification}~\cite{97-roTA,15-stocHS,16-stocHS,04-STL,07-robustMITL,09-robustSTL,13-monSTL}.
Delays on the other hand, are usually considered in system models~\cite{04-LazyRHA,17-roTA,08-Wulf,00-Puri,11-Channel}, which is the focus of this section.
In order to simplify presentation of later definitions, we next define a delay signal that assigns a delay to every pair of cars. 
If $\tau$ is a delay signal, its value for cars $c_1,c_2\oftype\CCC$ at time $t\oftype\nnReal$ models a delayed time in car $c_1$ when it observes
state of car $c_2$ at time $t$.
Although, we use one delay signal throughout our entire formulation, one can easily extend this to multiple delay signals, one for each part of the system.
\aref{def:delay} formally defines a delay signal based on response time parameter $\rho$.
Note that, by definition, there is no delay at time $t=0$.
Also, if $\rho=0$ then there will be no delay in the future either.

\begin{definition}[Delay Signal]\label{def:delay}
{\em Delay signal} is any piecewise continuous signal of type $\FTypeA{\nnReal}{\FTypeA{\CCC}{\FTypeA{\CCC}{\nnReal}}}$ that satisfies
$\forall t\oftype\nnReal,c_1,c_2\oftype\CCC\WeHave t-\rho\leq\tau\ t\ c_1\ c_2 \leq t$, where $\rho\oftype\nnReal$ is defined in \aref{tbl:params}. 
We use $\DELAY_\CCC$ to denote the set of all delay signals for cars in $\CCC$.
\end{definition}


The general idea in \cite{17-mob} to guarantee safety is to 
  first define a safe distance between every two cars and 
  then take a proper action whenever distance is unsafe.
The safe distance is computed using the knowledge a car has about velocity of itself and another car, and 
is {\em supposed to be} large enough such that the car will have enough time to respond properly, whenever the distance becomes unsafe.
\aref{def:dmin-lon} and \aref{def:dmin-lat} define minimum (safe) longitudinal and lateral distances, respectively.

\begin{definition}[Minimum Longitudinal Distance Signal]\label{def:dmin-lon}
Let $f$ and $\tau$ be, respectively, position and delay signals.
We define {\em minimum longitudinal distance signal}, denoted by $\dMinLon$, as a function of type
  $\FTypeA{\nnReal}{\FTypeA{\CCC}{\FTypeA{\CCC}{\pReal\cup\Brace{-\infty}}}}$ 
that maps 
  a time $t$ and 
  cars $c_1,c_2\oftype\CCC$ with
  $t_1\defEQ\tau\ t\ c_1\ c_1$ and
  $t_2\defEQ\tau\ t\ c_1\ c_2$ to
$-\infty$ if $c_1=c_2$ or $f\ t_1\ c_1\ y > f\ t_2\ c_2\ y$, and to the maximum of $\mu$ and 
following term, otherwise:
\begin{flalign*}
&  \rho\ f_v\ t_1\ c_1\ y      \ \ +\ \ 
   \frac{1}{2}\aMaxAccl\ \rho^2    \ \ +\ \  \\ 
&  \frac{\Paren*{f_v\ t_1\ c_1\ y \ +\  \rho\ \aMaxAccl}^2}{2\aMinBrake} \ \ -\ \ 
  \frac{\Paren*{f_v\ t_2\ c_2\ y}^2}{2\aMaxBrake}
\end{flalign*}
\end{definition}

There is a big difference between \aref{def:dmin-lon} and its correspondence in~\cite{17-mob}.
\aref{def:dmin-lon} uses delayed observations, but minimum distance in~\cite{17-mob} is defined assuming exact value of every car's longitudinal velocity
is known to every other car at all times.
Another difference is that, in \aref{def:dmin-lon}, we make sure minimum distance is never smaller than $\mu$, however in~\cite{17-mob} this distance can
become arbitrary close to $0$.
Since both \aref{def:dmin-lat} below and~\cite{17-mob} make sure that minimum lateral distance is never smaller than $\mu$, our approach is more uniform.

\begin{definition}[Minimum Lateral Distance Signal]\label{def:dmin-lat}
Let $f$ and $\tau$ be, respectively, position and delay signals.
We define {\em minimum lateral distance signal}, denoted by $\dMinLat$, as a function of type
  $\FTypeA{\nnReal}{\FTypeA{\CCC}{\FTypeA{\CCC}{\pReal\cup\Brace{-\infty}}}}$ 
that maps 
  a time $t$ and 
  cars $c_1,c_2\oftype\CCC$ with
  $t_1\defEQ\tau\ t\ c_1\ c_1$ and
  $t_2\defEQ\tau\ t\ c_1\ c_2$ to
$-\infty$ if $c_1=c_2$, and to the 
following term, otherwise:
\begin{flalign*}
& \mu + 
  \rho\Size*{f_v\ t_1\ c_1\ x}      \ \ +\ \ 
  \rho\Size*{f_v\ t_2\ c_2\ x}      \ \ +\ \ 
\\  
&\phantom{ \mu + \;\, }  
  \frac{1}{2}\aMaxAcclLat\ \rho^2   \ \ +\ \  
  \frac{\Paren*{\Size{f_v\ t_1\ c_1\ x} \ +\  \rho\ \aMaxAcclLat}^2}{2\aMinBrakeLat}\ \ +\\
&\phantom{ \mu + \;\, }  
  \frac{1}{2}\aMaxAcclLat\ \rho^2   \ \ +\ \  
  \frac{\Paren*{\Size*{f_v\ t_2\ c_2\ x} \ +\  \rho\ \aMaxAcclLat}^2}{2\aMinBrakeLat}
\end{flalign*}
\end{definition}

Note that similar to \aref{def:dmin-lon}, minimum distance in \aref{def:dmin-lat} is also computed using delayed observations.
The only other difference between \aref{def:dmin-lat} and its correspondence in~\cite{17-mob} is that we do not assume car $c_1$ is on the left of car $c_2$.
Finally, to the best of our knowledge, the case when two cars move laterally in the same direction is not considered in~\cite{17-mob} and hence nor
here~\footnote{%
We leave it to the reader to prove or disprove the necessity of considering that case.}.

According to \aref{def:pos-sig} and what comes after it, for any position signal $f$, there are unique velocity and acceleration signals.
However, according to \aref{def:dmin-lon} and \aref{def:dmin-lat}, 
when response time ($\rho$) is positive, there could be uncountably many minimum longitudinal/lateral distance signals for $f$.
This is because, we assume velocity and acceleration are physical properties that are defined using position.
For example, if positions at times $1$ and $3$ are respectively $10$ and $18$ then (average) velocity during this time is exactly $\frac{18-10}{3-1}$.
However, we assume actual values of these signals are obtained/observed with delay of at most $\rho$ units of time.
Signal $f$ (and hence signals $f_v$ and $f_a$) can take uncountably many values during any positive duration of time.
Therefore, there are uncountably many possible minimum longitudinal/lateral distance signals that can be observed/considered.


\begin{definition}[Dangerous Longitudinal Situation Signal]\label{def:dang-lon}
Let $f$, $\tau$, and $\dMinLon$ be a 
  position, 
  delay, and
  minimum longitudinal distance signals, respectively.
We define {\em dangerous longitudinal situation signal}, denoted by $\DangLon$, as a function of type
  $\FTypeA{\nnReal}{\FTypeA{\CCC}{\FTypeA{\CCC}{\Brace{\top,\bot}}}}$
that maps 
  a time $t$ and 
  cars $c_1,c_2\oftype\CCC$ with
  $t_1\defEQ\tau\ t\ c_1\ c_1$ and
  $t_2\defEQ\tau\ t\ c_1\ c_2$ to
to $\top$ exactly when
value of $\dMinLon\ t\ c_1\ c_2$ is strictly larger than 
$\Paren{f\ t_2\ c_2\ y} - \Paren{f\ t_1\ c_1\ y}$.
\end{definition}

\begin{definition}[Dangerous Lateral Situation Signal]\label{def:dang-lat}
Let $f$, $\tau$, and $\dMinLat$ be a 
  position, 
  delay, and
  minimum lateral distance signals, respectively.
We define {\em dangerous lateral situation signal}, denoted by $\DangLat$, as a function of type
  $\FTypeA{\nnReal}{\FTypeA{\CCC}{\FTypeA{\CCC}{\Brace{\top,\bot}}}}$
that maps 
  a time $t$ and 
  cars $c_1,c_2\oftype\CCC$ with
  $t_1\defEQ\tau\ t\ c_1\ c_1$ and
  $t_2\defEQ\tau\ t\ c_1\ c_2$ to
to $\top$ exactly when
value of $\dMinLat\ t\ c_1\ c_2$ is strictly larger than 
$\Size{\Paren{f\ t_2\ c_2\ x} - \Paren{f\ t_1\ c_1\ x}}$.
\end{definition}

We define {\em dangerous situation signal}, denoted by $\Dang$, as a function of type
  $\FTypeA{\nnReal}{\FTypeA{\CCC}{\FTypeA{\CCC}{\Brace{\top,\bot}}}}$
that maps time $t\oftype\nnReal$ and cars $c_1,c_2\oftype\CCC$ to
the conjunction of $\Paren{\DangLon\ t\ c_1\ c_2}$ and $\Paren{\DangLat\ t\ c_1\ c_2}$.
Note that \aref{def:dang-lon} and \aref{def:dang-lat} also use delayed observations in their definitions.


\begin{definition}[Blame Time Signal]\label{def:blame}
Let $\Dang$ be a dangerous situation signal.
We define {\em blame time signal}, denoted by $\Blame$, as a function of type
  $\FTypeA{\nnReal}{\FTypeA{\CCC}{\FTypeA{\CCC}{\nnReal\cup\Brace{\infty}}}}$
that maps 
  time $t\oftype\nnReal$ and
  cars $c_1,c_2\oftype\CCC$ 
  to\\ 
\begin{tabular}{l@{\ \ \ if\ \ \ }l}
$\infty$  & $\neg\Dang\ t\ c_1\ c_2$ or $\forall r\oftype\LclRop{0,t}\WeHave\Dang\ t\ c_1\ c_2$~\footnotemark\\
$t'$      & \begin{tabular}[t]{@{}l@{}}
            $\forall r\oftype\LopRcl{t',t}\WeHave\Dang\ r\ c_1\ c_2$ and \\
            $\forall t''\oftype\LclRop{0,t'}\WeHave\exists r'\oftype\LopRcl{t'',t'}\SuchThat\neg\Dang\ r'\ c_1\ c_2$ \\
            \end{tabular}
\end{tabular}
\footnotetext{%
      The case $\forall r\oftype\LclRop{0,t}\WeHave\Dang\ t\ c_1\ c_2$ is only considered here for completeness.
      However, using additional constraints that will be given later, we would not consider any signal that is initially dangerous, 
      \ie\ satisfies $\exists c_1,c_2\oftype\CCC\SuchThat\Dang\ 0\ c_1\ c_2$
      (note that value of $\Dang\ 0\ c_1\ c_2$ is uniquely determined by values of $f\ 0\ c_1\ c_2$ and $f_v\ 0\ c_1\ c_2$).}
\\
We denote the blame time signals that are obtained by replacing $\Dang$ with $\DangLon$ and $\DangLat$, respectively by $\BlameLon$ and $\BlameLat$.
\end{definition}

The second condition in \aref{def:blame} uniquely defines value of $t'$.
Intuitively, it is the smallest value $t'$ for which the situation is dangerous at any time between $t'$ and $t$.
Note that according to~\cite{17-mob}, in the second case of \aref{def:blame}, instead of 
$\forall t''\oftype\LclRop{0,t'}\WeHave\exists r'\oftype\LopRcl{t'',t'}\SuchThat\neg\Dang\ r'\ c_1\ c_2$, 
we should have just said $\neg\Dang\ t'\ c_1\ c_2$.
It is easy to see that our condition is strictly weaker.
For example, if $t'>0$, $\Dang\ t'\ c_1\ c_2=\top$ and $\forall r'\oftype\Opened{0,t'}\WeHave\neg\Dang\ r'\ c_1\ c_2$ then
value of $\Blame\ t\ c_1\ c_2$, according to \aref{def:blame} is $t'$, and according to \cite{17-mob} is undefined.
We leave it to the reader to (dis)prove that \aref{def:dang-lon} and \aref{def:dang-lat} and whatever comes before them guarantee dangerous  
signal is continuous from left, in which case \aref{def:blame} and its correspondent in~\cite{17-mob} are equivalent.


According to the following quote from~\cite{17-mob}, by simply not moving, a car can have $0$ longitudinal velocity for a long time, but it is impossible
for a car to keep its lateral velocity at $0$. We believe this is a mistake, since if a car does not move then it has zero velocity in both 
directions.
Furthermore, since we only consider position signals with fully defined velocity and acceleration, the velocity signal is continuous throughout the entire time.
Therefore, whenever its sign is different at time $t_1$ and $t_2$, we know its value is $0$ at some time between $t_1$ and $t_2$, which is enough for the purpose of
this paper.
\begin{quote}
Unlike longitudinal velocity, which can be kept to a value of 0 for a long time (the car is simply not moving), 
keeping lateral velocity at exact $0$ is impossible as cars usually perform small lateral fluctuations. 
It is therefore required to introduce a robust notion of lateral velocity.
\end{quote}


We have everything we need to finally define a policy in \aref{def:policy}.

\begin{definition}[Policy]\label{def:policy}
Let $f$ and $\tau$ be position and delay signals, respectively, and 
let signals $\dMinLon$, $\dMinLat$, $\DangLon$, $\DangLat$, $\Dang$, $\BlameLon$, $\BlameLat$, and $\Blame$ be uniquely defined 
based on $f$ and $\tau$, as specified in this section.
For any time $t\oftype\nnReal$ and car $c_1\oftype\CCC$,
we say car $c_1$ follows the policy at time $t$, denoted by $\Policy\ f\ \tau\ t\ c_1$ \iFF\
for any car $c_2\oftype\CCC$,
  if $\Dang\ t\ c_1\ c_2=\top$ and $t_b\defEQ\Blame\ t\ c_1\ c_2\in\Real$ then
  the following conditions hold: 
\begin{itemize}
\item If before the blame time there was a safe longitudinal distance between $c_1$ and $c_2$
      (\ie\ $t_b=\BlameLon\ t\ c_1\ c_2$) then
      \begin{enumerate}
      \item $\forall t'\oftype\Opened{t_b,t_b+\rho}\WeHave f_a\ t'\ c_1\ y \leq \aMaxAccl$, \ie\
            within the response time, acceleration of the rear car must be bounded by \aMaxAccl.
      \item $\forall t'\oftype\Closed{t_b+\rho,t}\WeHave f_a\ t'\ c_1\ y \leq -\aMinBrake$, \ie\
            after the response time, bound on acceleration decreases to $-\aMinBrake$
            (the rear car must use its brake).
      \item $\forall t'\oftype\LopRcl{t_b,t}\WeHave f_a\ t'\ c_2\ y \geq -\aMaxBrake$, \ie\
            there is bound on how fast the front car can stop.
      \end{enumerate}
\item If before the blame time there was a safe lateral distance between $c_1$ and $c_2$
      (\ie\ $t_b=\BlameLat\ t\ c_1\ c_2$) then
      \begin{enumerate}
      \item $\forall t'\oftype\Opened{t_b,t_b+\rho}\WeHave \Size{f_a\ t'\ c_1\ x} \leq \aMaxAcclLat$, \ie\
            within the response time, acceleration of car $c_1$ must be bounded by \aMaxAcclLat.
      \item $\forall t'\oftype\Closed{t_b+\rho,t}\WeHave \Size{f_a\ t'\ c_1\ x} \leq \aMinBrakeLat$ and $(f_v\ t'\ c_1\ x)\times(f_a\ t'\ c_1\ x)\leq0$, \ie\
            after the response time, bound on acceleration decreases to $\aMinBrakeLat$ and acceleration and velocities are in the opposite direction.
            (\aka\ the $c_1$ must use its brake).
      \end{enumerate}
\end{itemize}
We define $\Policy\ f\ \tau\ c_1$ to be $\forall t\oftype\nnReal\WeHave\Policy\ f\ \tau\ t\ c_1$ (\ie\ car $c_1$ always follows the policy). Similarly,
we define $\Policy\ f\ \tau$ to be $\forall c_1\oftype\CCC\WeHave\Policy\ f\ \tau\ c_1$ (\ie\ every car follows the policty at all time).
\end{definition}

There are three differences between the first part of policy written in \aref{def:policy} and the one introduced in~\cite{17-mob}.
First, according to \aref{def:policy}, there is no requirement on acceleration of the rear car at time $t_b$.
We believe imposing a restriction at time $t_b$ is a mistake, specially in~\cite{17-mob}, since by definition the situation is not dangerous at $t_b$ and no car
can look into the future of the system state.
The next two differences are more important.
According to~\cite{17-mob}, after the rear car reached to full stop, it can never move forward.
Similarly, after the front  car reached to full stop it can never decelerate.
We believe either one of these policies is too restrictive to be allowed in any real scenario.
One implies if the rear car enters a dangerous situation, it is going to stop on the road and never move again.
The other one implies if the front car enter into a dangerous situation with a car on its behind, first it will fully stop and then if it moves, it will never 
lower its speed. None of these makes any sense in practical scenarios.
These three differences also exists between the second part of policy written in \aref{def:policy} and the one introduced in~\cite{17-mob}.

\section{Verification Problems}\label{sec:probs}

We have specified dynamics and policy of cars in \aref{sec:spec}.
In this section we specify multiple verification problems about those specifications.
According to \aref{sec:spec}, 
for every position and delay signals, 
minimum longitudinal and lateral distance signals (\dMinLon\ and \dMinLat),
longitudinal and lateral dangerous situation signals (\DangLon, \DangLat, and \Dang), and
longitudinal and lateral blame time signals (\BlameLon, \BlameLon, and \Blame), are all uniquely defined.
Therefore, in this section, whenever we consider a position and a delay signal, we assume all the other signals can be used without introduction.
We divide our verification problems into three different categories:
\begin{inparaenum}
\item safety         properties,
\item liveness       properties, and
\item responsibility properties.
\end{inparaenum}

\subsection{Safety Problems}\label{sec:safety}

\begin{problem}[Safety]\label{prb:safe}
Prove or disprove that policy  in \aref{def:policy} guarantees utopia (\ie\ prevents accident).
More precisely, prove or disprove the following formula cannot be satisfied by a position signal $f\oftype\LOC_\CCC$:
\begin{gather*}
  \overbrace{
  \Paren*{
      \exists\tau\oftype\DELAY_\CCC  \WeHave    
      \raisebox{0pt}[\height][10pt]{
        \ensuremath{
          \underbrace{\Paren*{\forall c_1,c_2\oftype\CCC\WeHave\neg\Dang\ 0\ c_1\ c_2}}_{\AAA_2}
        }
      }
  \wedge \Policy\ f\ \tau} }^{\AAA_1}
  \wedge\\[0.6\baselineskip]
  \underbrace{
  \exists t\oftype\nnReal,c_1,c_2\oftype\CCC\WeHave f\ t\ c_1 = f\ t\ c_2}_{\AAA_3}
\end{gather*}
\end{problem}
Condition $\AAA_2$ guarantees that the situation is not initially dangerous.
Condition $\AAA_1$ guarantees $f$ is initially not dangerous and it follows policy as specified in \aref{def:policy}.
Condition $\AAA_3$ guarantees that there will be an accident in the future.
A system/policy is {\em safe} \iFF\ the formula defined in \aref{prb:safe} is unsatisfiable.
Finally note that formula defined in \aref{prb:safe} depends on parameters given in \aref{tbl:params}. 
We leave it to the reader to solve this problem for only one or a class of values of these parameters.

\aref{prb:safe} ultimately depends on signals $\dMinLat$ and $\dMinLon$.
Distances defined by these two signals are never smaller than the same distances defined in~\cite{17-mob}.
However, as we mentioned multiple times, there is a big difference here: in this paper 
definitions of signals $\dMinLat$ and $\dMinLon$ involve delay, while in~\cite{17-mob} these signals are defined 
using no delay (\ie\ response time is zero)~\footnote{%
  Authors in~\cite{17-mob} only consider positive response time when a car responds to a dangerous situation.
  However, observing position and velocity of the every other car that is used to determine if a situation is dangerous 
  is assumed to be performed within $0$ response time.}.
Furthermore, minimum/safe distance defined in \aref{def:dmin-lon} uses the fact that observations are made with
no delay. This intuitively means that, using policy and minimum distance defined in~\cite{17-mob}, cars can become
arbitrary close to each other.
Therefore, it should be of no surprise that if we compute minimum distance the same way as in~\cite{17-mob}, but use delayed 
values for it, cars will crash.
This informal justification answers \aref{prb:safe} negatively.
However, it is not clear to us how one should fix this problem. 
For example, if we consider delay, is it still true that the minimum distance is always exists, or even to guarantee its 
existence one has to bound both velocity and acceleration (policy in \aref{def:policy} only bounds signals during some 
intervals)?
Furthermore, validity of any suggestion for fixing this issue requires a formal proof,
something that we look forward to be done automatically.

\aref{prb:safe} completely ignores errors and uncertainties in each cars' sensors.
As mentioned before, this is usually handled in the context of robust verification. 
Note that there are many definitions for robustness. What we put here is taken from~\cite{97-roTA} and is for illustration purposes only.
For any position signal $f\oftype\LOC_\CCC$ and $\epsilon\oftype\nnReal$, let $\infBall^\epsilon(f)$ be the set of signals in $\LOC_\CCC$ 
that are point-wise $\epsilon$-close to $f$. More precisely, 
$f'\in\infBall^\epsilon(f)$ \iFF\ $\sup_{t\oftype\nnReal}d(f\ t, f'\ t) \leq \epsilon$, where $d$ is the distance function used in the 
definition of signal.

\begin{definition}[$\epsilon$-Robust Safe and Unsafe Signals]\label{def:rob-safe}
Let $\AAA_1$ and $\AAA_3$ be the two predicates over position signals defined in \aref{prb:safe}.
A position signal 
$f\oftype\LOC_\CCC$ is called {\rm $\epsilon$-robust safe} \iFF\ it satisfies the following formula:
\[
  \AAA_1(f) \Rightarrow \forall f'\oftype\infBall^{\epsilon}(f)\WeHave \neg\AAA_3(f')
\]
Similarly, 
$f\oftype\LOC_\CCC$ is called {\em $\epsilon$-robust unsafe} \iFF\ it satisfies the following formula:
\[
  \AAA_1(f) \wedge \forall f'\oftype\infBall^{\epsilon}(f)\WeHave\AAA_3(f')
\]
A position signal is called {\em robustly safe (unsafe)} \iFF\ it is $\epsilon$-robust safe (unsafe) for some $\epsilon\oftype\pReal$.
A policy is called {\em $\epsilon$-robust safe (unsafe)} \iFF\ all (some) position signals are $\epsilon$-robust safe (unsafe) in that policy.
A policy is called {\em robustly safe (unsafe)} \iFF\ all (some) position signals are robustly safe (unsafe) in that policy.
\end{definition}
Note that it is impossible for a position signal (or a policy) to be both robustly safe and unsafe. But
it is possible for a position signal (or a policy) to be neither robustly safe nor robustly unsafe.

\begin{problem}[Robust Safety]\label{prb:rob-safe}
Prove or disprove that the policy in \aref{def:policy} is robustly safe (or robustly unsafe).  
More precisely, determine which of the following sentences are true and which ones are false:
\begin{enumerate}
\item $\epsilon$-robust safe:\ \ \ \ \ 
      $\forall f\oftype\LOC_\CCC\WeHave
      \AAA_1(f) \Rightarrow \forall f'\oftype\infBall^{\epsilon}(f)\WeHave \neg\AAA_3(f')$
\item $\epsilon$-robust unsafe:
      $\exists f\oftype\LOC_\CCC\SuchThat
      \AAA_1(f) \wedge \forall f'\oftype\infBall^{\epsilon}(f)\WeHave\AAA_3(f')$
\item robustly safe:\\
      $\forall f\oftype\LOC_\CCC\WeHave\exists\epsilon\oftype\pReal\SuchThat
      \AAA_1(f) \Rightarrow \forall f'\oftype\infBall^{\epsilon}(f)\WeHave \neg\AAA_3(f')$
\item robustly unsafe:\\
      $\exists f\oftype\LOC_\CCC\WeHave\exists\epsilon\oftype\pReal\SuchThat
      \AAA_1(f) \wedge \forall f'\oftype\infBall^{\epsilon}(f)\WeHave\AAA_3(f')$
\end{enumerate}
\end{problem}
It should be easy to see that being $\epsilon$-robust safe (unsafe) implies being robustly safe (unsafe). But the converse is not necessarily true.

\subsection{Liveness Problems}\label{sec:liveness}

Having a safe system is not enough, otherwise one could write \texttt{false} as the simplest policy that guarantees safety of every system 
(\ie\ any behavior that satisfies this policy is safe). 
We also need to make sure that it is possible for a signal to satisfy the policy that is specified in \aref{def:policy}.

\begin{problem}[Liveness]\label{prb:live}
Prove or disprove that policy in \aref{def:policy} is not inconsistent (\ie\ it can be followed).
More precisely, prove or disprove the following formula can be satisfied by a position signal $f\oftype\LOC_\CCC$:
\begin{gather*}
  \forall\tau\oftype\DELAY_\CCC  \SuchThat    
  \Paren{\forall c_1,c_2\oftype\CCC\WeHave\neg\Dang\ 0\ c_1\ c_2} \wedge \Policy\ f\ \tau
\end{gather*}
\end{problem}
In \aref{prb:safe} we have $\exists\tau\oftype\DELAY_\CCC$, but 
in \aref{prb:live}   we have $\forall\tau\oftype\DELAY_\CCC$. 
We chose to have this change, since 
  for safety   we want to say using any valid delay signal that together with the position signal follow the policy, results in a safe behavior. However, 
  for liveness we want to say there is a position signal for which {\em any} delay signal can be used to follow the policy.
This is because the intended use of delay is to allow a bounded amount of response time, and any amount of delay within this bound should be allowed by the 
policy.

Similar to the case of \aref{prb:safe} vs.\ \aref{prb:live}, just having liveness is not enough.
Otherwise, although in theory there is a signal $f\oftype\LOC_\CCC$ that follows the policy, 
in practice, a car has to always behave exactly like $f$ (any deviation violates the policy) which is never possible.

\begin{definition}[$\epsilon$-Robust Live Signals]\label{def:rob-safe}
A position signal 
$f\oftype\LOC_\CCC$ is called {\em $\epsilon$-robust live} \iFF\ it satisfies the following formula:
\[
  \forall f'\oftype\infBall^{\epsilon}(f) \WeHave
  \text{$f'$ satisfies the formula defined in \aref{prb:live}}  
\]
A position signal is called {\em robustly live} \iFF\ it is $\epsilon$-robust live for some $\epsilon\oftype\pReal$.
A policy is called {\em robustly live} \iFF\ some position signal $f\oftype\POS_\CCC$ is robustly live in it.
\end{definition}

\begin{problem}[Robust Liveness]\label{prb:rob-long-live}
Prove or disprove that the policy in \aref{def:policy} is robustly live.
\end{problem}

Our definition of (robust) liveness is the minimum requirement for system to be considered live and in practice one has to add more constraints to it.
For example, in order to consider a position signal $f$ live, one might want to also consider the following two constraints.
\begin{inparaenum}
\item Longitudinal position of every car diverges to infinity ($\forall u\oftype\Real,c\oftype\CCC\WeHave\exists t\oftype\nnReal\SuchThat f\ t\ c\ y > u$). 
			Otherwise, a policy that does not move any car will be considered $\infty$-robust safe and live.
\item There are always points in time at which all cars are moving for a positive duration of time
			($\forall t\oftype\nnReal\WeHave
				\exists t_1\oftype\Opened{t,\infty},t_2\oftype\Opened{t_1,\infty}\SuchThat
				\forall r\oftype\Opened{t_1,t_2},c\oftype\CCC\WeHave 
				f_v\ t\ c\ y>0$).
			Otherwise, a policy that moves only one car at a time can be considered robustly live.
\end{inparaenum}
{\em Determining} the exact set of constraints for liveness is {\em not} a formal process and should be determined using experience or simulation.


\subsection{Responsibility-Sensitive Safety Problem}\label{sec:rss-probs}

Our problems in \aref{sec:safety} and \aref{sec:liveness} only concern the case in which every car follows the policy.
However, there is always someone on the road who drives recklessly. 
Authors in~\cite{17-mob}, introduce the concept of ``who is responsible for an accident'', and 
instead of trying to come up with a policy that guarantees absence of an accident, they come of with a policy that guarantees if a car follows the policy then it 
won't be held responsible for an accident.

\begin{definition}[Responsibility for an Accident]\label{def:rss}
Let $f$ and $\tau$ be position and delay signals, respectively.
Let $c_1,c_2\oftype\CCC$ be two cars, and 
let $t\oftype\nnReal$ be a time of accident between $c_1$ and $c_2$ (\ie\ $f\ t\ c_1 = f\ t\ c_2$).
We say $c_1$ is {\em responsible for the accident with $c_2$ at time $t$} \iFF\ 
$\Dang\ t\ c_1\ c_2=\top$ and $c_1$ did not follow the policy (as specified in \aref{def:policy}) at sometime during $\LopRcl{t_b,t}$, where 
$t_b\defEQ\Blame\ t\ c_1\ c_2$.
\end{definition}

Once again considering delays distinguishes \aref{def:rss} from the same definition in~\cite{17-mob}.
For example, because of delays, blame time ($t_b$) for an accident could be different in $c_1$ and $c_2$.
Even worse, it is not so much obvious that whenever there is an accident, there will be a blame time.
We consider these problems next.
However, it should be obvious that according to \aref{def:rss}, whoever follows the policy won't be held responsible for an accident.

\begin{theorem}[Responsibility-Sensitive Safety]\label{thm:rss}
Whoever follows the policy won't be held responsible for an accident.
\end{theorem}


\begin{problem}[Existence of Responsible Party]\label{prb:rss}
Prove or disprove that each accident has at least one responsible party.
More precisely, prove the following formula cannot be satisfied by any signal $f\oftype\LOC_\CCC$.
\begin{gather*}
  \exists
    \tau           \oftype\DELAY_\CCC,
    c_1,c_2        \oftype\CCC,
    t              \oftype\nnReal,
    t_{b_1},t_{b_2}\oftype\nnReal\cup\Brace{\infty}
  \SuchThat\\
  f\ t\ c_1 = c\ t\ c_2   \ \wedge\ 
  \neg \Dang\ 0\ c_1\ c_2 \ \wedge\ 
  \neg \Dang\ 0\ c_2\ c_1 \\
  \Paren*{t_{b_1}=\Blame\ t\ c_1\ c_2   \ \ \wedge\ \ \forall t'\oftype\LopRcl{t_{b_1},t}\cap\nnReal\WeHave\Policy\ f\ t'\ \tau\ c_1} \wedge\\
  \Paren*{t_{b_2}=\Blame\ t\ c_2\ c_1   \ \ \wedge\ \ \forall t'\oftype\LopRcl{t_{b_2},t}\cap\nnReal\WeHave\Policy\ f\ t'\ \tau\ c_2} \phantom{\wedge}
\end{gather*}
\end{problem}



\section{Tools}\label{sec:tools}
In \aref{sec:spec} and \aref{sec:probs}, we defined system specifications as well as five different fundamental problems about those specifications.
In this section, we look at different formal verification tools, and for each tool we specify why our problems cannot be even expressed using
these tools. All of these tools are developed solely for the purpose of model checking cyber-physical systems. 
\aref{tbl:tools} lists these tools along where they fail to support required features.
We have identified six reasons.
The first four prevent us from specifying our models using these tools, and the last two prevents us from specifying our verification problems using 
these tools.
Note that we completely ignored 
  possible difficulties in expressing our models and problems in the language of these tools, and
  the fact that \CCC, finite set of cars, is given as a parameter (\ie\ it is fine if a tool can solve these problems for a fixed known number of cars $\geq2$).

\NewDocumentCommand{\HAREx}{}{\HARE\textsubscript{16}}
\NewDocumentCommand{\HAREv}{}{\HARE\textsubscript{17}}

\begin{itemize}
\item Non-Linear Dynamics: 
      Some tools do not support non-linear dynamics.
      For example, 
        \UPPALL\ is for model checking \TAlp,
        \HyTech\ is for model checking \RHAlp,
        \SpaceEx, \PHAVer, and \HAREx~\footnote{%
          \HAREx~\cite{16-HARE} and \HAREv~\cite{17-HARE} are two different versions of the same tool.
          We decided to separate them since only the older version supports ordinary differential inclusion.} 
        are for model checking hybrid automata with affine dynamics.
      Note that support for non-linear dynamics in \HAREv, is only for flows and not discrete transitions.
\item Ordinary Differential Inclusions (ODI):
      Some tools only support ordinary differential equation and not ordinary differential inclusion.
      In \aref{sec:spec}, the only constraints that we ever put on accelerations was some bound on its value in \aref{def:policy}.
      This means velocity is restricted using some bound on its derivative.
\item Delays in Dynamics:
      None of these tools supports having delays in dynamics.
      In \TAlp, delays in dynamics are closely related to skewed clocks, and for a very large subclass of \TAlp, it is known how to handle skew clocks 
      using \UPPALL~\cite{17-roTA,08-Wulf,00-Puri,11-Channel}. However, \TAlp\ are far from what we need to express our dynamics.
      Note that even if we set response time ($\rho$) to $0$, blame time and hence policy still depend on continuous state of the system in the past.
\item Unbounded State Space:
      Most tools that handle non-linear dynamics, require state space to be bounded using intervals for every state variable.
      However, no state variable is bounded in this paper.
\item Unbounded Time:
      Similar to unbounded state space, most tools that handle non-linear dynamics require time horizon to be bounded.
      Note that bounding time does not necessarily bound number of discrete transitions that can be taken within the given bound~\cite{13-TBRfMHA,14-LnReach},
      and tools like \dReach, \CtEt, and \FlowS\ also require number of discrete transitions to be bounded as well.
\item Robustness:   
      None of these tools supports specifying robustness.
      Similar to delays, in \TAlp, robustness (as defined in this paper) is similar to perturbing constraints.
      If we consider perturbation of constraints for robustness, the problem has been already solved for timed automata using 
      \UPPALL~\cite{17-roTA,08-Wulf,00-Puri,11-Channel}. However, not only \TAls, is far from what we need in our specification, it is not even clear that 
      robustness as defined here (taken from~\cite{97-roTA}) is equivalent to robustness based on perturbation of constraints.
\end{itemize}
 
\begin{table}[H]
\NewDocumentCommand{\xmark}{}{\ding{55}}
\NewDocumentCommand{\Head}{}{\color{white}}
\NewDocumentCommand{\ROT}{m}{\rotatebox[origin=c]{90}{\ #1\ }}
\begin{tabular}{|ll|*{6}{c|}}
\rowcolor{black}
\Head Name           &  &
\Head \ROT{Non-}\ROT{Linearity}  & 
\Head \ROT{ODI}            & 
\Head \ROT{Delays}         & 
\Head \ROT{Unbounded} \ROT{State} &
\Head \ROT{Unbounded} \ROT{Time}&
\Head \ROT{Robustness} \\
\dReach            &\cite{15-dReach}  &      &\xmark&\xmark&\xmark&\xmark&\xmark\\\hline
\SpaceEx           &\cite{11-SpaceEx} &\xmark&\xmark&\xmark&\xmark&\xmark&\xmark\\\hline
\PHAVer            &\cite{05-PHAVer}  &\xmark&      &\xmark&      &      &\xmark\\\hline
\HyTech            &\cite{97-HyTech}  &\xmark&      &\xmark&      &      &\xmark\\\hline
\CtEt              &\cite{15-C2E2}    &      &\xmark&\xmark&\xmark&\xmark&\xmark\\\hline
\FlowS             &\cite{13-FlowStar}&      &\xmark&\xmark&\xmark&\xmark&\xmark\\\hline
\HAREx             &\cite{16-HARE}    &\xmark&      &\xmark&      &      &\xmark\\\hline
\HAREv             &\cite{17-HARE}    &      &\xmark&\xmark&      &      &\xmark\\\hline
\HSolver           &\cite{07-HSolver} &      &      &\xmark&\xmark&      &\xmark\\\hline
\UPPALL            &\cite{uppaal}     &\xmark&\xmark&\xmark&      &      &\xmark\\\hline
\end{tabular}
\caption{%
  Different model checkers and why they cannot be used to solve our problems.
  Cross marks are where a tool lacks a required support.
}\label{tbl:tools}
\end{table} 

\section{Conclusion}\label{sec:conc}

In this paper, we presented a challenge problem for formal verification tools developed or aimed to be developed for industrial cyber-physical system.
We formalized main components of dynamics and policies introduced in~\cite{17-mob} for autonomous vehicles driving on a straight road. 
This also helped us to find some inconsistencies with the current specifications in~\cite{17-mob}.
To the best of our knowledge, no current automatic formal verification tool can be used to even express these dynamics and problems.
We hope this serves as a challenge problem for formal tools targeting automatic verification of industrial cyber-physical systems.

\bibliographystyle{ACM-Reference-Format}
\bibliography{biblio}

\end{document}